\documentclass[fleqn,usenatbib]{mnras}

\usepackage{newtxtext,newtxmath}
\usepackage[dvipsnames]{xcolor}
\definecolor{xlinkcolor}{cmyk}{1,1,0,0}
\usepackage[T1]{fontenc}

\DeclareRobustCommand{\VAN}[3]{#2}
\let\VANthebibliography\thebibliography
\def\thebibliography{\DeclareRobustCommand{\VAN}[3]{##3}\VANthebibliography}

\usepackage{comment}
\usepackage{graphicx}	% Including figure files
\usepackage{amsmath}	% Advanced maths commands
\usepackage{xspace}	
\usepackage{lineno}
\usepackage{natbib}
\usepackage{xcolor}
\usepackage{multicol}
\usepackage{pdflscape}

\newcommand{\todo}[1]{\textbf{\color{red} #1}}

\newcommand{\asat}{{\em AstroSat}\xspace}
\newcommand{\fermi}{{\em Fermi}\xspace}
\newcommand{\kw}{{\em Konus}-Wind\xspace}

\newcommand{\swift}{{\em Swift}\xspace}

\newcommand{\mm}{{Mass Model}\xspace}

\title{Localisation of Gamma Ray Bursts using \asat~\mm}

\author[Saraogi et al.]{Divita Saraogi$^{1},$\thanks{E-mail: divitadsaraogi@gmail.com}
J Venkata Aditya$^{2}$, 
Varun Bhalerao$^{1},$\thanks{E-mail: varunb@iitb.ac.in}
Suman Bala$^{1,3}$, 
Arvind Balasubramanian$^{4}$, 
\newauthor
Sujay Mate$^{4}$, 
Tanmoy Chattopadhyay$^{5}$, 
Soumya Gupta$^{6}$, 
Vipul Prasad$^{8}$,
Gaurav Waratkar$^{1}$,
\newauthor
Navaneeth P K$^{8}$,
Rahul Gopalakrishnan $^{8}$,
Dipankar Bhattacharya$^{7}$, 
Gulab Dewangan$^{8}$, 
Santosh Vadawale$^{9}$
\\
$^{1}$Department of Physics, Indian Institute of Technology Bombay, Powai, Mumbai, Maharashtra  400076, India\\
$^{2}$Department of Computer Science and Engineering, Indian Institute of Technology Bombay, Powai, Mumbai, Maharashtra 400076, India\\
$^{3}$Science and Technology Institute, Universities Space Research Association, Huntsville, AL 35805, USA\\
$^{4}$Department of Astronomy and Astrophysics, Tata Institute of Fundamental Research, Mumbai, Maharashtra  400005, India\\
$^{5}$Kavli Institute of Astrophysics and Cosmology, Stanford University, 452 Lomita Mall, Stanford, CA 94305, USA\\
$^{6}$Homi Bhabha National Institute, Anushakti Nagar, Mumbai, Maharashtra 400094, India\\
$^{7}$Ashoka University, Department of Physics, Sonepat, Haryana 131029, India\\
$^{8}$Inter University Centre for Astronomy and Astrophysics, Pune, Maharashtra 411007, India\\
$^{9}$Physical Research Laboratory, Ahmedabad, Gujarat 380009, India\\
}

\date{Accepted XXX. Received YYY; in original form ZZZ}

\pubyear{2023}
\begin{document}
\label{firstpage}
\pagerange{\pageref{firstpage}--\pageref{lastpage}}
\maketitle

\begin{abstract}
The Cadmium Zinc Telluride Imager (CZTI) aboard \asat has good sensitivity to Gamma Ray Bursts (GRBs), with close to 600 detections including about 50 discoveries undetected by other missions. However, CZTI was not designed to be a GRB monitor and lacks localisation capabilities. We introduce a new method of localising GRBs using `shadows' cast on the CZTI detector plane due to absorption and scattering by satellite components and instruments. Comparing the observed distribution of counts on the detector plane with simulated distributions with the \asat~\mm, we can localise GRBs in the sky. Our localisation uncertainty is defined by a two-component model, with a narrow Gaussian component that has close to 50\% probability of containing the source, and the remaining spread over a broader Gaussian component with an 11.3 times higher $\sigma$. The width ($\sigma$) of the Gaussian components scales inversely with source counts. We test this model by applying the method to GRBs with known positions and find good agreement between the model and observations. This new ability expands the utility of CZTI in the study of GRBs and other rapid high-energy transients.
%We studied the localisation capabilities of Astrosat using the data of blah GRBs which have reference locations from other GRB detection instruments. The results show that AstroSat is capable of localising GRBs with a systematic error of about $\blah^{\circ}$. For typical GRB in 70 keV to 200 keV with a fluence from 10−5 erg/cm 2 to 10 −4 erg/cm 2, the statistical error of localisation is between $\blah^{\circ}$ to $\blah^{\circ}$ . 
%We studied the localisation capabilities of Astrosat using the data of blah GRBs which have reference locations from other GRB detection instruments.The results show that AstroSat is capable to localise GRBs with a systematic error of about $\blah^{\circ}$. For typical GRB in 70 keV to 200 keV with a fluence from 10−5 erg/cm 2 to 10 −4 erg/cm 2 , the statistical error of localisation is between $\blah^{\circ}$ to $\blah^{\circ}$ . 
\end{abstract}


%%%%%%%%%%%%%%%%%%%% BODY OF PAPER %%%%%%%%%%%%%%%%%%%%%%%%%%%%%%%%%%%%%%%

\section{Introduction}
%\subsection{HE Transients (GRBs +EMGW+SGRs+etc)}
%\subsection{Localisation}
%Why is localisation important --- position for follow up etc --- matters a lot for rare events \\
%typical localisation methods --- coded mask (bat) --- projection (gbm, gecam) --- ipn
%\subsection{Earlier attempts and current work}
%gw170817 exclusion by EO --- 170105 with raytrace code --- here we extend to using detailed sim called mass model
%-----------------------------------------------------------------------------------------------------------------------------------------------------------------------------------------------------
High-energy transients such as Gamma-Ray Bursts (GRBs), soft gamma repeaters (SGRs), X-ray counterparts to fast radio bursts (FRBs), and electromagnetic counterparts to gravitational waves (EMGW) are extraordinary physical phenomena linked to the activities of compact objects like neutron stars and black holes. Recent years have seen increased interest in the search for and study of these high-energy transients due to their association with cosmological sources \citep{2019NatRP...1..585M}. Among these phenomena, GRBs stand out as particularly intriguing transient events, releasing an astonishing amount of energy of the order of $10^{51}$ to $10^{54}$ erg \citep{piron2016gamma}. The initial prompt emission during a GRB occurs within a brief duration, typically ranging from a fraction of a second to a few minutes --- or even thousands of seconds in the longest cases \citep{zhang2014gamma,2020ApJ...893...46V,2016ApJ...829....7L}. However, the brevity and uncertain positional information of these events make their study challenging.

GRBs have been a subject of study for several decades. To track the afterglow of GRBs, numerous ground-based telescope networks have been established, complementing space-based instruments that initially detect the prompt emission for initial localisation. Survey telescopes like Panoramic Survey Telescope and Rapid Response System \citep[PAN-STARRS;][]{chambers2016pan} and the Zwicky Transient Facility \citep{bellm2014zwicky} diligently follow up on transients and relay transient alerts through Astronomers Telegram \citep{hlovzek2019data}. Additionally, global radio telescopes, including Very Long Baseline Interferometry \citep[VLBI;][]{chandra2016gamma,mundell2010gamma}, conduct follow-up studies of these transients. By studying transients across different wavelength bands, a comprehensive understanding of the GRB emission mechanism can be achieved, aiding the development and refinement of theoretical models.

%Rare events are key : 170817: -BAT, CZTI, (GW170104 & GRB170105 - Tale of two transients), 2 IPN GRBs.
The detection of GRB~170817A \textemdash~the gamma ray burst associated with the gravitational wave event GW170817 along with the gravitational wave detection by Laser Interferometer Gravitational-wave Observatory (LIGO) has given birth to multi-messenger astronomy \citep{abbott2017gravitational}. The detection of this GRB by \fermi \citep{goldstein2017ordinary} coincident with the gravitational wave signal spurred a flurry of multi-wavelength observations that led to detailed characterisation of this event. While \fermi \citep{goldstein2017ordinary} and \emph{Integral} \citep{2017ApJ...848L..15S} detected this burst, it was missed by \swift\ and \asat\ which were sensitive enough to detect it as it was occulted by the Earth \citep{2017Sci...358.1565E,2017Sci...358.1559K}. The non-detection by CZTI despite the flux being above the detection threshold proved the Earth-occulted scenario and aided in narrowing down the source localisation by a factor of two \citep{2017Sci...358.1559K}. Detection and localisation of high-energy bursts has at times changed the complete interpretation of an event. The discovery of GRB~170105A and its localisation were key in proving that the orphan afterglow ATLAS17aeu was associated with this GRB and not the binary black hole merger GW170104 \citep{bhalerao2017tale}. \asat-CZTI has been detecting a large number of bursts and can be used to measure the polarisation of bright bursts \citep{2019ApJ...884..123C,chattopadhyay2022hard}. However, polarisation analysis requires the knowledge of the source position in the sky, failing which the analysis cannot be completed. Two examples of such bursts are GRB~200503A \citep{gupta2020astrosat} and GRB~201009A \citep{gupta2020grb}. While they were also detected by other spacecraft like \textit{AGILE} \citep{ursi2020grb_1,ursi2020grb}, \kw (Hurley, private communication), the sources were not well-localised by any of them. This has prevented us thus far from making two valuable additions to the rather small number of GRBs with measured polarisation.

%Types of localisations 
Astronomical X-ray instruments employ various methods for source localisation. One approach involves projection-based localisation, where the number of counts observed by different detectors of the same instrument facing different directions is compared. This technique is used by missions such as \fermi-GBM \citep{connaughton2015localization}, \textit{CGRO}-BATSE \citep{meegan1992spatial}, \textit{GECAM} \citep{xin2020localization}, and the upcoming \textit{Daksha} mission \citep{bhalerao2022daksha}, providing localisations within a few degrees. Another method utilizes a coded aperture mask (CAM) to achieve enhanced localisation accuracy of the order of a few arc minutes. Instruments like \swift-BAT \citep{barthelmy2005burst}, \textit{INTEGRAL} \citep{mereghetti2004real}, and ECLAIRS onboard the \textit{SVOM} mission \citep{triou2009eclairs} utilize CAM for GRB localisation, but this improved localisation typically comes at the cost of a limited field of view. Additionally, the interplanetary network (IPN) technique, involving data from different satellites \citep{hurley2013interplanetary,2020ApJ...905...82H}, enables GRB localisation by leveraging the separation and arrival time of bursts on multiple satellites.

%earlier attempts and current work
To localise GRBs with \asat-CZTI, earlier attempts have been made by using ray-trace simulations which is an analytical method calculating the shadow (i.e. the degree of absorption and transmission) cast by different components of CZTI on the detector, for photons of given energy and direction of incidence and predict the effective area of the same \citep{bhalerao2017tale}. However, this method ignores the absorption or scattering contributions of the rest of the satellite, and hence is rather limited in its accuracy and the fraction of the sky where it is applicable. Hence, to localise the GRBs we present a method that uses the shadow cast by other instruments, allowing us to determine the GRB location by comparing the observations with simulations obtained using the \asat~\mm \citep{mate2021astrosat}. In this paper, we present the localisation results obtained by analyzing a sample of GRBs detected by \asat-CZTI, using the \asat~\mm to localise these GRBs and validate our results using localisation information from other instruments like \fermi, \swift or Interplanetary network (IPN).

%\rough{the rough direction of the GRB is readily known from the intensity of the GRB in various quadrants, and from visual inspection of the detector plane histograms. To refine this further, we had to do more work - which we present in this paper.}
 This paper is structured in the following fashion: We briefly describe the \asat-CZTI instrument and the \mm in section \S\ref{sec:trans_CZTI}. The framework for localising  GRBs using the \asat~\mm is discussed in section \S\ref{sec:loc_principle}. The localisation results for our GRB sample and the calculation of uncertainty on the result are discussed in section \S\ref{sec:results}. In section \S\ref{sec:conclusion}, we summarize our results and describe the impact of our results on future work.
%=====================================================================================================================================================================================================
\section{Transients with \asat-CZTI}\label{sec:trans_CZTI}
%what is czti --- how many grbs --- etc\\
%Study many HE TRANSIENTS (cite -CIFT, Polarisation sample paper, Akash’s FRB paper(maybe))
%\subsection{\asat Mass Model}
%\begin{enumerate}
%    \item what is mass model?
%    \item Concept of DPH
%    \item How mass model can be utilised for localisation
%\end{enumerate}

\asat is India's first multi-wavelength mission that carries instruments covering the energy range from ultraviolet to high-energy X-rays \citep{singh2014astrosat}. The five principal scientific payloads onboard \asat are: (i) a Soft X-ray Telescope (SXT), (ii) three Large Area Xenon Proportional Counters (LAXPCs), (iii) a Cadmium-Zinc-Telluride Imager (CZTI), (iv) an Ultra-Violet Imaging Telescope (UVIT) configured as two independent telescopes, and (v) a Scanning Sky Monitor (SSM). The CZTI instrument onboard \asat is a high-energy X-ray instrument providing wide-field coverage. CZTI contains 64 detectors with 256 pixels each, arranged in a square pattern to obtain nearly 1000~cm$^2$ geometric area \citep{bhalerao2017cadmium}. Collimators limit the primary field of view to approximately 4.6\degr, but these become increasingly transparent to radiation above 100~keV. This, combined with the 20--200~keV sensitive energy range makes CZTI an excellent all-sky detector for high-energy transients. CZTI was not designed as a GRB detection instrument, but over the years of its operation, it has proven very successful in detecting a large number of GRBs \citep{sharma2021search}. In the 7 years of its operation \asat has detected close to 600 GRBs\footnote{\url{http://astrosat.iucaa.in/czti}}. \asat has also proven very successful in measuring the polarisation of GRBs, having studied 20 GRBs so far \citep{chattopadhyay2022hard}.

As \asat can detect off-axis GRBs but does not have localisation capability, it has thus far relied on other GRB detection instruments for localisation information. The localisation method presented here can overcome this hurdle.\\

\begin{figure}
\includegraphics[width=0.5\textwidth]{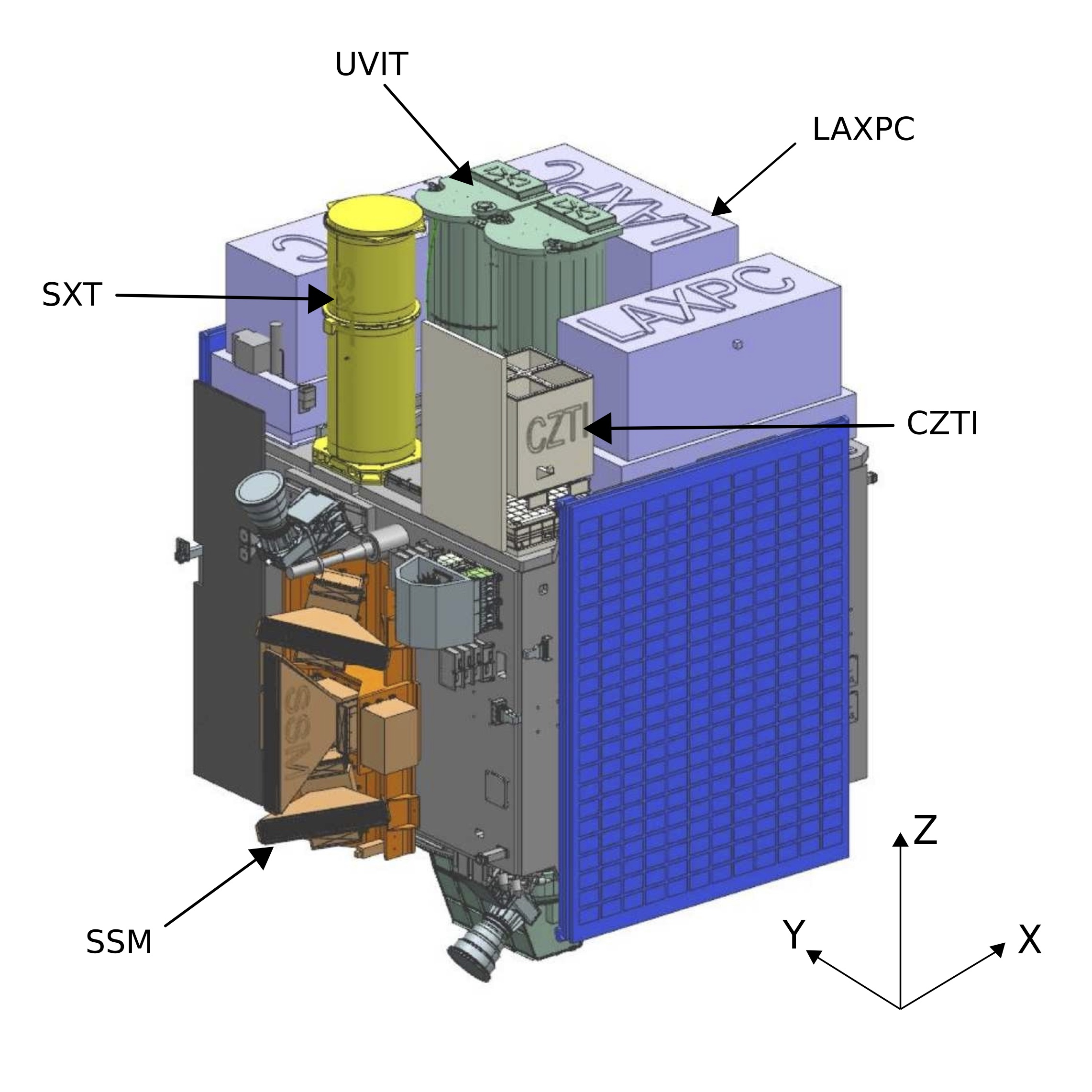}
\caption{Representative CAD model of \asat\ with folded solar panels, featuring the five payloads of \asat, and displaying the coordinate axes for CZTI. \label{fig:astrosat}}
\end{figure}
% \citep[See for instance][and references therein]{somepaper}
%\textbf{\asat Mass Model}\\
The \asat Mass Model is a numerical model of the satellite body, created using GEANT4\footnote{\url{http://geant4.web.cern.ch/geant4/}} \citep{agostinelli2003geant4}, a tool for simulating interactions of photons and particles with matter. The details of \asat Mass Model can be found in \citet{mate2021astrosat}, hereafter referred to as MM21. We use the Mass Model to simulate the interaction of the incoming GRB photons with the satellite elements. Photons from a source incident on the satellite body may get transmitted, scattered, or absorbed and re-emitted with new energies and directions. All these effects lead to a direction-dependent `image' on the detectors, which we term a `Detector Plane Histogram' (DPH). Numerical calculations with the mass model can be used to simulate this DPH as a function of source direction. Thanks to the inhomogeneous distribution of satellite elements around the CZTI detectors, the DPH is different for each source direction. Thus, we can use the observed DPH for any GRB and compare it with a library of simulated DPHs to infer the location of the source in the sky. This principle is used to localise GRBs. 
%========================================================================================================
\section{Localisation using Mass Model}\label{sec:loc_principle}
%\subsection{How shadows cased by other instruments can be used to localise GRBs}
%\begin{enumerate}
%    \item how shadows change with source location
%    \item chisquare minimization
%    \item scaling correction - why? improvement?
%\end{enumerate}

%\subsection{How we localise in practice}
%\begin{enumerate}
%    \item HTM Grid - what? why? grid points
%    \item simulations
%    \item how many energies, for how many grid points
%   \item grid point spacing
 %   \item how grid was created
 %   \item Utilisation of IUCAA cluster
 %   \item time required for generating the whole grid
%\end{enumerate}

%\subsection{Localisation process}
%\begin{enumerate}
%    \item Chisq comparison
%    \item contour plots
%    \item what binning is applied
%    \item what binning is used for representation (in DPH)
%    \item effect of scaling
%\end{enumerate}

\begin{figure}
\includegraphics[width=0.5\textwidth]{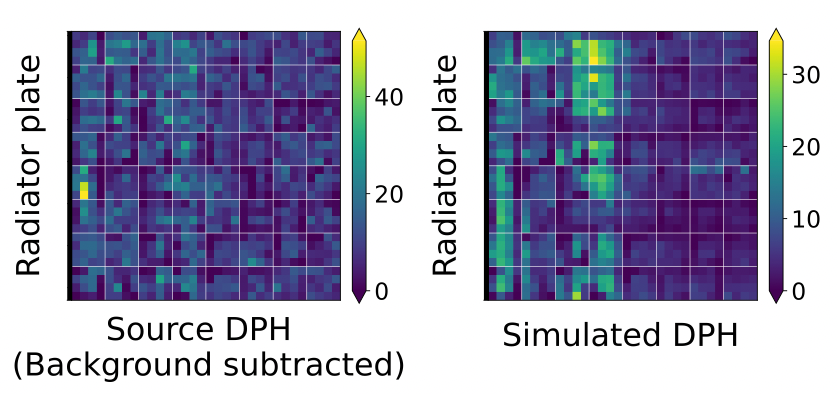}%{plots/GRB190117A_T054.00_P177.48_70.0_200.0-pages.pdf}
\caption{Comparing detector plane histograms (DPHs) for GRB~190117A. The left figure shows the observed background-subtracted DPH. The right side shows the scaled simulated DPH for a source located at the closest HTM grid point to the actual GRB position. The left image is more ``grainy'' due to the small number of statistics in observed data.\label{fig:one_dph}}
\end{figure}

%what is a DPH and how DPH pattern changes with position
In our localisation approach, we employ simulations generated by the \asat~\mm at various positions in the sky. These simulations yield a DPH giving the expected number of photons in each of the 16,384 pixels of CZTI. 
%As expected, these DPHs change with the source position, as illustrated in Figure~\ref{fig:dph}.
This simulated DPH can then be compared with the observed background-subtracted DPH of a GRB. 
We bin the pixel data to perform a module-wise comparison between the observed data to the simulated DPH, as given by Equation~\ref{eq:chisq}:
\begin{equation}
    \chi^2 = \sum_{m=1}^{64} \frac{(N_\mathrm{sim,m} - N_\mathrm{src,m})^2}{\sigma_{sim,m}^2 + \sigma_{src,m}^2} \label{eq:chisq}
\end{equation}
where the summation is carried out over all 64 modules and on an average each module has more than 60 counts. The uncertainties in the simulated DPH are calculated as described in MM21. The GRB window $t_\mathrm{src}$ contains $N_\mathrm{obs,m}$ counts from the GRB as well as background. We calculate the source counts and uncertainty as follows:

\begin{eqnarray}
    N_\mathrm{src,m} & = & N_\mathrm{obs,m} - N_\mathrm{bkg,m} \times \frac{t_\mathrm{src}}{t_\mathrm{bkg}} \\
    \sigma_{src,m}^2 & = & N_\mathrm{obs,m} + N_\mathrm{bkg,m} \times \bigg(\frac{t_\mathrm{src}}{t_\mathrm{bkg}}\bigg)^{2}
\end{eqnarray}
The background $N_\mathrm{bkg,m}$ is measured using a longer window $t_\mathrm{bkg}$ and scaled before subtracting, in order to reduce the uncertainty contribution. The uncertainties in source and background are calculated assuming Poisson statistics.
Repeating this procedure over the entire sky, the global minimum of $\chi^2$ gives the best-fit position of the source. As discussed in MM21, in practice it has been seen that there can be an overall scaling factor needed to make the GRB fluence match with the observed counts, and an offset factor to correct for any background subtraction residuals. Following MM21, we allow for additive and multiplicative scaling factors while calculating the $\chi^2$ values for each direction in our grid. As an example, Figure~\ref{fig:one_dph} shows the source DPH for GRB~190117A, compared with our simulated DPH for the closest grid point to the true GRB location on the sky.

%In practice how do we localise GRBs. %We take known GRBs to validate localisation capability
For localisation, we need simulations uniformly distributed over the entire sky. The selection of such points can be conveniently done using grids like the Hierarchical Triangular Mesh (HTM) \citep{kunszt2001hierarchical} or Healpix \citep{gorski2005healpix}. For this work, HTM offers a distinct advantage: as we refine the resolution of HTM pixels, the centres of pixels from a coarse grid also become points on a finer grid. This allows us to refine our grid in selected parts of the sky without disrupting the grid pattern. Healpix, on the other hand, does not share this property. Hence, we opt to forgo the Healpix grid and instead pick an HTM grid of an appropriate resolution. %We found that an HTM ``Level 4'' grid with an inter-pixel spacing of 5.62\degr\ is an appropriate choice for our work. %We found that a HTM ``Level 4'' grid with an inter-pixel spacing of 5.62\degr\ is an appropriate choice for our work. 

%The Hierarchical Triangular Mesh grid (HTM) operates by defining a point within a two-dimensional coordinate system on a spherical surface. This grid involves subdividing the surface into nearly-equivalent triangles, each assigned a distinct identifier for locating objects within a containing polygon. Utilizing the HTM grid offers the added benefit of obviating the need to simulate a large number of points during grid refinement, as the center of the level 4 grid remains consistent with the center of the level 5 grid.
%The level 4 HTM grid has been created using mass model simulations for GRB with flat spectrum. For each directions there are 197 energy simulations as described in the MM21. The overall sky area is divided almost evenly into 2048 grid points. The separation between these points is $5.62^{\circ}$. For simulating these points we used the IUCAA cluster using 16 nodes each with 8 cores for a total of $\sim 10^{5}$ CPU-hours.
Based on simulations, we concluded that we need a grid resolution of $\sim 5^{\circ}$. This corresponds to a `Level 4' HTM grid with pixel centres separated by $5.62^{\circ}$, having a total of 2048 pixels in the sky. We generated the Level 4 HTM grid in satellite $\theta$--$\phi$ coordinates, where $\theta$ is measured from the $Z$-axis and $\phi$ is measured from the $X$-axis along the $XY$ plane (Figure \ref{fig:astrosat}). We then undertook GEANT4 \mm simulations at each of those points. Since each GRB can have a different spectrum, we opted to simulate DPHs for monochromatic incident photons at 197 energies ranging from 20~keV -- 2~MeV, with an adaptive step size based on energy resolution. The simulations commence at 20 keV, with increments of 5 keV up to 500 keV. In the range of 500 keV to 1 MeV, the step size is increased to 10 keV, followed by 20 keV steps in the 1 MeV to 2 MeV range. As discussed in MM21, photons up to about 2~MeV can make significant contributions in the 20--200~keV range. At each energy level, we simulate the interaction of $6.97$ million incident photons with the entire satellite. This simulation is carried out by employing a circular source plane with a radius of $200$ cm, resulting in a photon flux of $55.46$ photons cm$^{-2}$ as outlined in MM21. To carry out these simulations, we harnessed the high-performance computing cluster at IUCAA, Pune\footnote{\url{http://hpc.iucaa.in/}}, leveraging 16 nodes, each equipped with 512 cores, which collectively accumulated to approximately $10^{5}$ CPU-hours.

The DPHs are then weighted by the spectrum of the incident GRB and summed to create the final simulated DPH for that direction. In cases where the GRB spectrum is unknown, we proceed with standard \fermi\ spectral parameters from \citet{2011A&A...530A..21N}. Once the location is known, in principle, the mass model can be used to fit the GRB spectrum, which can be used iteratively to improve the location. Such iterative or joint fitting is beyond the scope of this work and will be addressed in a separate future work.

% In future work we will try to address joint spectral and location fitting. 
% Once the DPHs are obtained, we compare them with the observations and calculate the $\chi^{2}$ for each module following the procedure discussed in detail in MM21. As discussed in detail there, we have sometimes noticed an overall scaling where an offset accounts for erroneous background subtraction and slope accounts for flux mismatches. For a given observed GRB, this procedure is used to calculate a $\chi^{2}$ for every simulated direction, we use this to create $\chi^{2}$ contours in $\theta-\phi$ space

As an example, we show the $\chi^2$ contours for GRB~190117A in Figure \ref{fig:all_sky_contour}. The best-fit position is then simply taken as the grid point having the lowest $\chi^{2}$ value. We note that the finite grid size means that even for the brightest GRB our $\theta$ offset will be as large as half of the grid size i.e. about $2.81^{\circ}$ or about the area of a pixel which is $20.14~\mathrm{deg}^{2}$.
%=====================================================SECTION4======================================================
\section{Method}\label{sec:results}
\subsection{GRB sample}
To validate our localisation results we select a sample of 29 bright GRBs detected by \asat before May 2023 and compare the results with the \mm simulations. The selection criterion for this sample is that the total GRB counts in the data should be $\ge 4000$ and the localisation information should be available from other missions like \fermi, \swift, \textit{Maxi}, and IPN so that we can validate our localisation results. We converted the RA, and Dec obtained from these instruments to the $\theta - \phi$ coordinates in the CZTI frame. For these GRBs, the spectral parameters are obtained from either \fermi or \kw, and we scale the normalization values in the energy range from $70-200$ keV, which is the energy range in which we perform all our analyses for \asat-CZTI. One source we do not consider in our sample is GRB~160325A which was detected with 19229 counts. It was incident nearly on-axis for CZTI, where accurate localisation is done using the coded aperture mask \citep{2021JApA...42...76V}. On the other hand, the \mm makes certain approximations which make it relatively less accurate for on-axis sources. Hence, we exclude this GRB from our sample.

\subsection{Localisation contours}\label{sec:contours}
%grid, chisq , maps, chisq reduces, we cannot directly use delta chisq
%\rough{For all the GRBs in the sample, we create $chi^{2}$ contours which gives us the localisation area for the GRB. We apply scaling correction to the data and use the minimum scaled chisq to get the grid position nearest to which the GRB is located according to the \mm simulations. During this calculations we remove the grid points which fall under the area which is occulted by the earth.}

% For the GRBs in our sample, we compare the Source DPH with the simulated DPH for all the 2048 grid points. We compute $\chi^{2}$ after applying scaling correction and find the grid point with the minimum $\chi^{2}$ value. Ideally, this point should be closest to the actual GRB location. We plot contours which give us the region of highest probability where the source is located. When calculating the minimum $\chi^{2}$ value to determine the source location, we eliminate the grid points that were occulted by the Earth during the particular orbit when the GRB was detected. Figure \ref{fig:all_sky_contour} shows the all-sky contour plot for GRB~190117A in RA-Dec and $\theta-\phi$ coordinate system.

\begin{figure*}
        \centering
            %\usebox{\bigpicturebox}\hfill
            %\begin{minipage}[b][\ht\bigpicturebox][s]{.45\textwidth}
            % TRIM: left bottom right top
            %\subfigure[RA=0, Dec=-90]
            {\includegraphics[trim=10 10 10 10,clip=true,width=.3\textwidth]{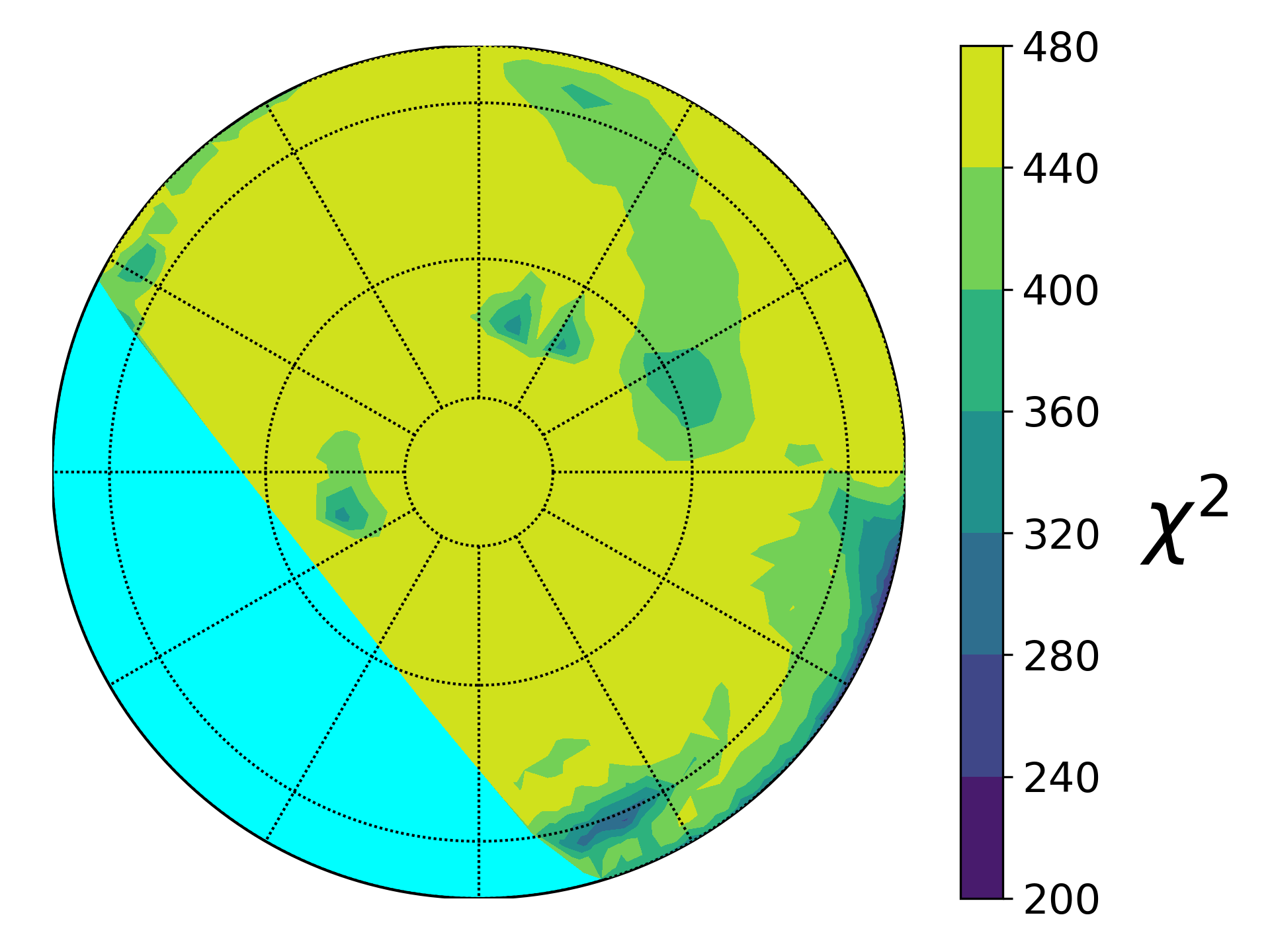}}\hspace{2em}
            {\includegraphics[trim=10 10 10 10,clip=true,width=.3\textwidth]{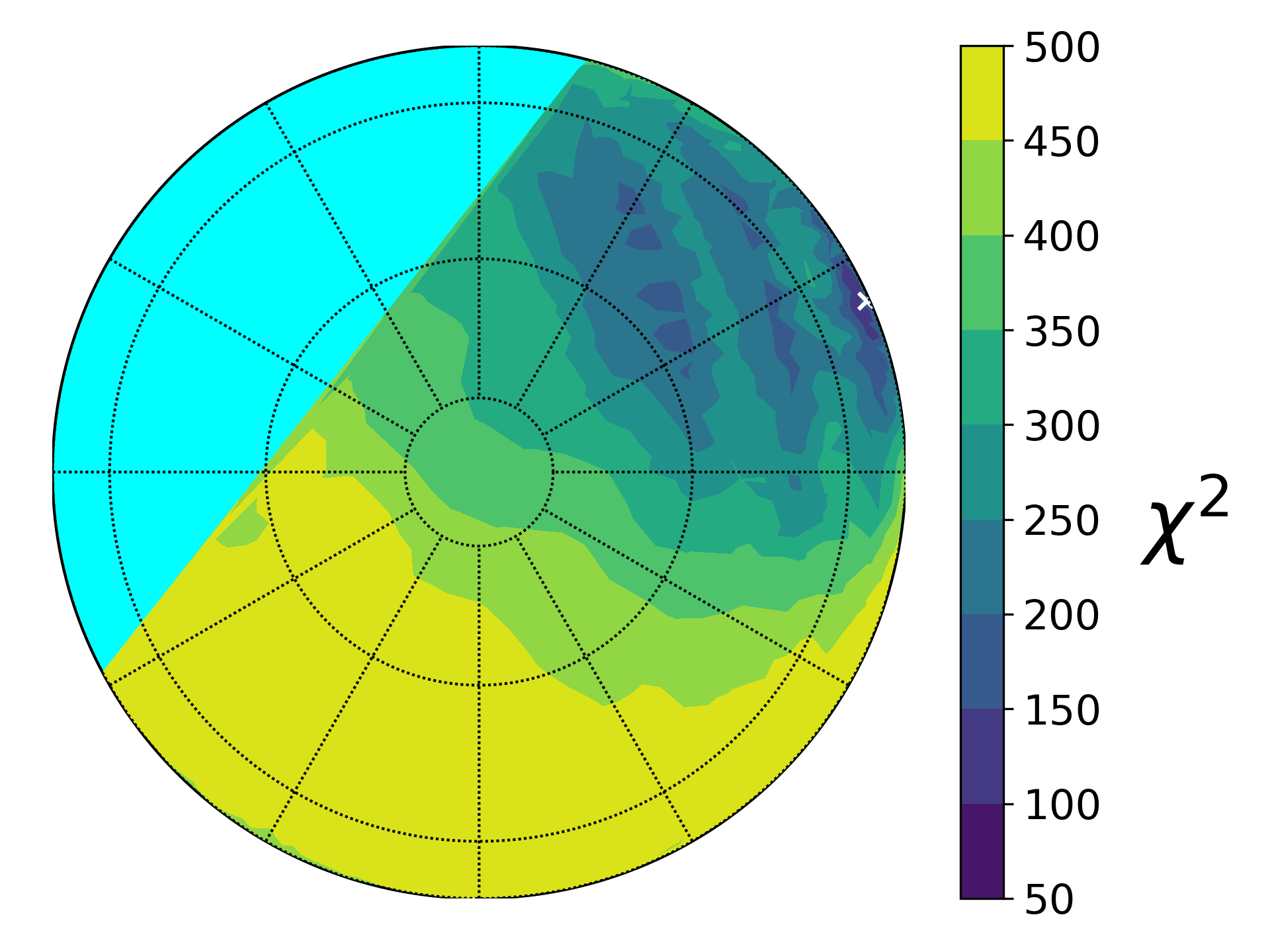}}\hspace{2em}
            {\includegraphics[trim=10 10 10 10,clip=true,width=.3\textwidth]{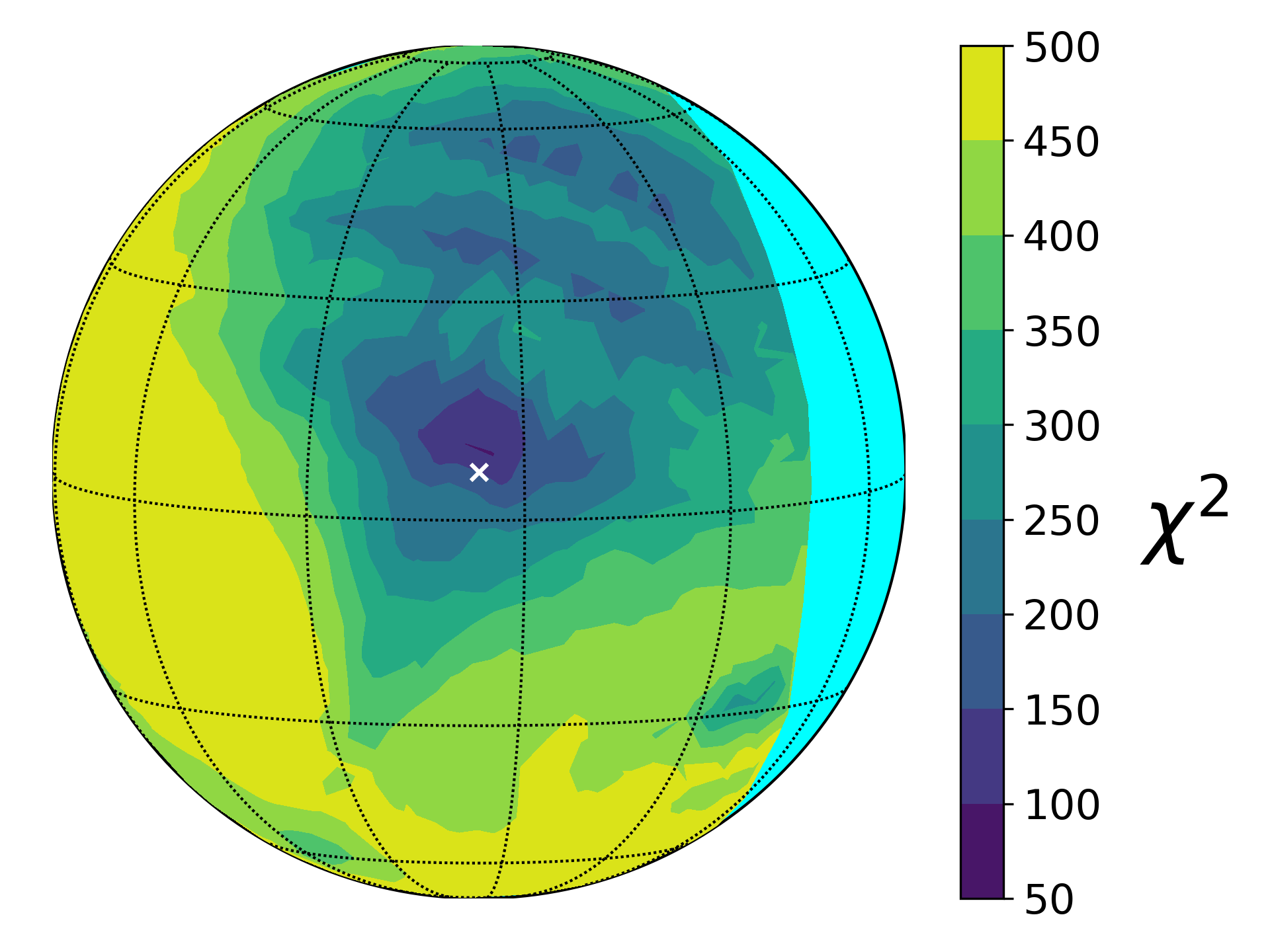}}\hspace{2em}\vfill
            %{\subfigure[$\theta=0, \phi=0$]
            {\includegraphics[trim=10 10 10 10,clip=true,width=.3\textwidth]{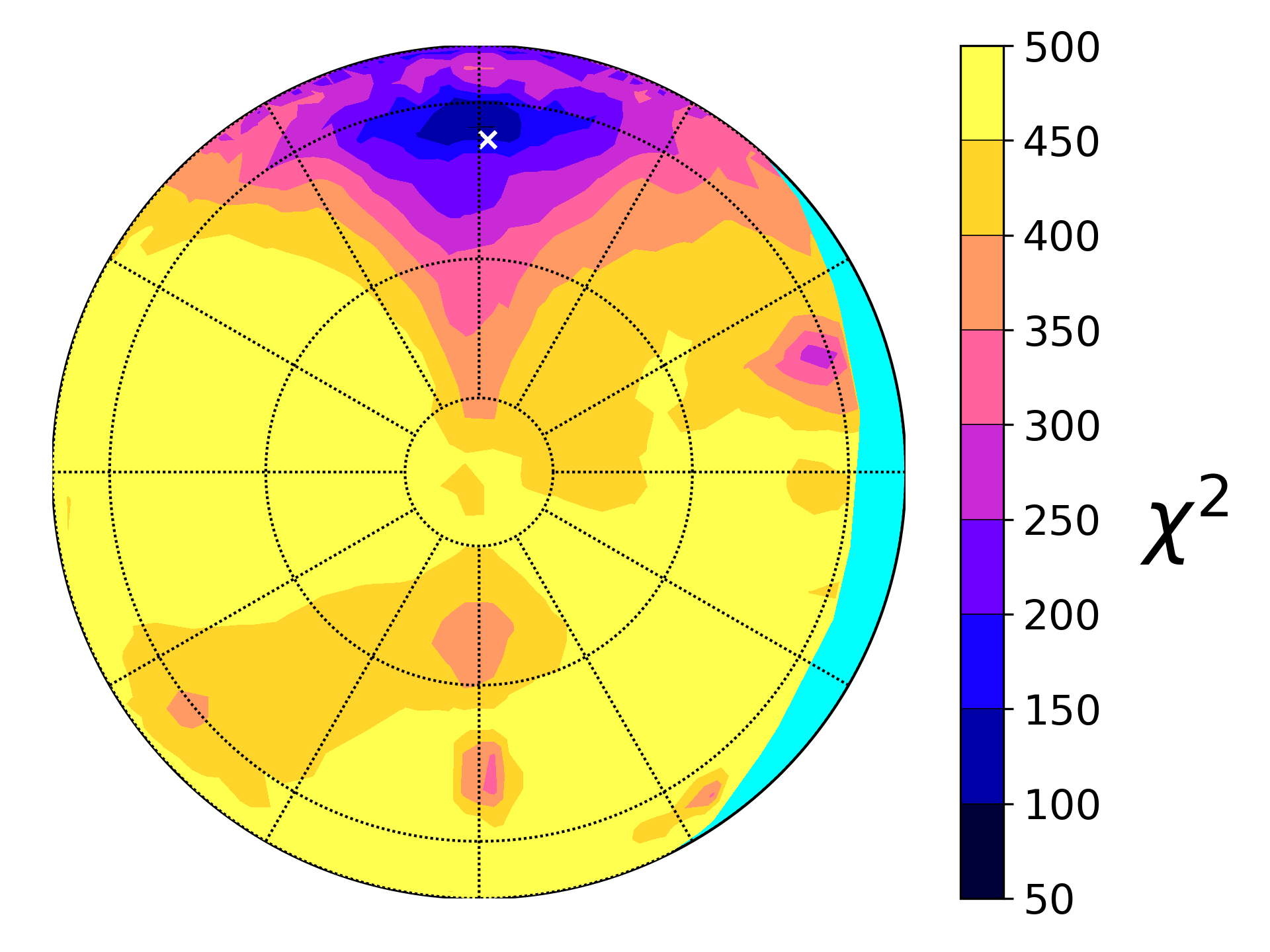}}\hspace{2em}
            %\subfigure[RA=0, Dec=90]
            %\subfigure[$\theta=180, \phi=0$]
            {\includegraphics[trim=10 10 10 10,clip=true,width=.3\textwidth]{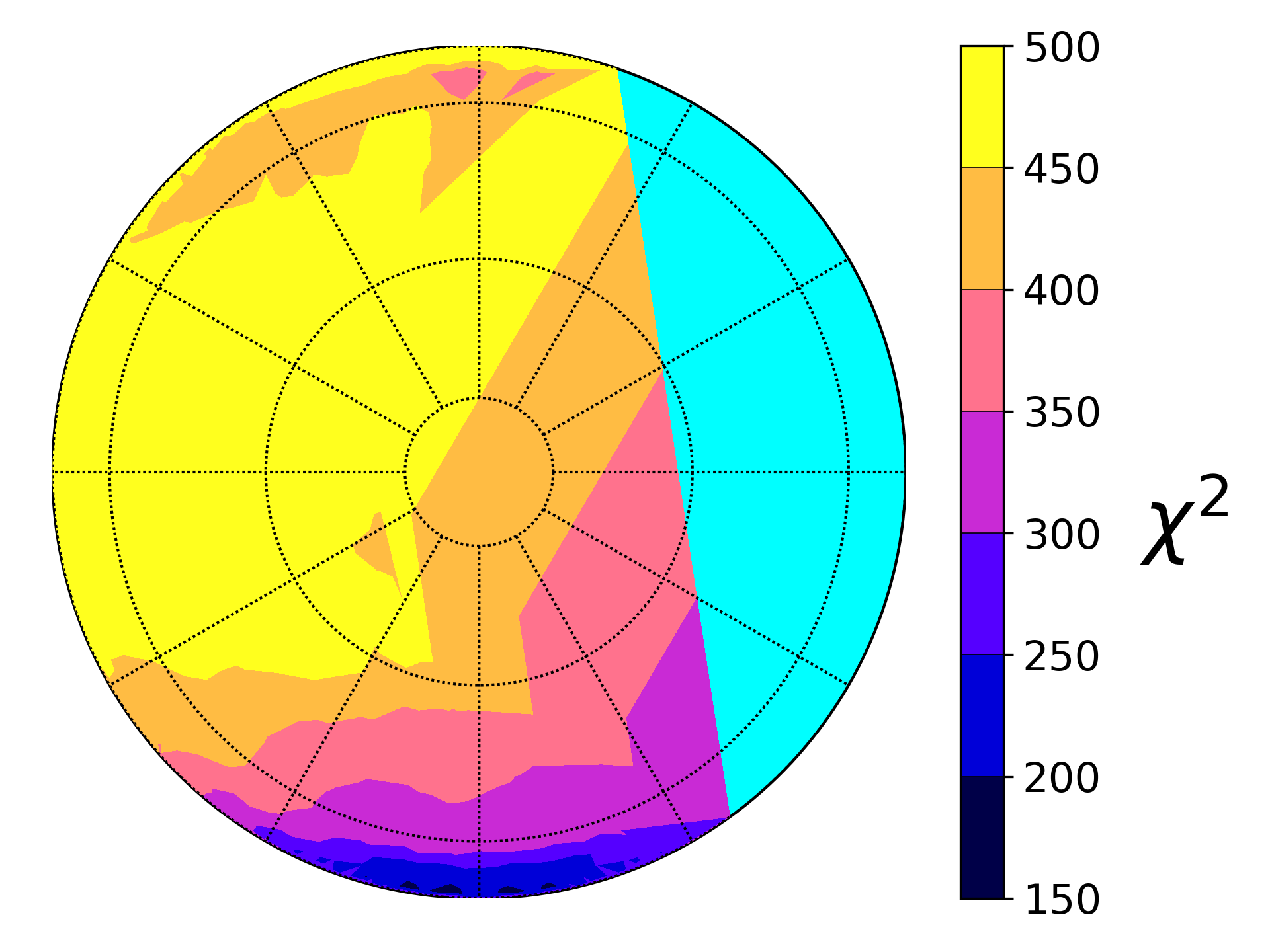}}\hspace{2em}
            %\subfigure[RA=113.86, Dec=6.51]
            %\subfigure[$\theta=51.26, \phi=178.51$]
            {\includegraphics[trim=10 10 10 10,clip=true,width=.3\textwidth]{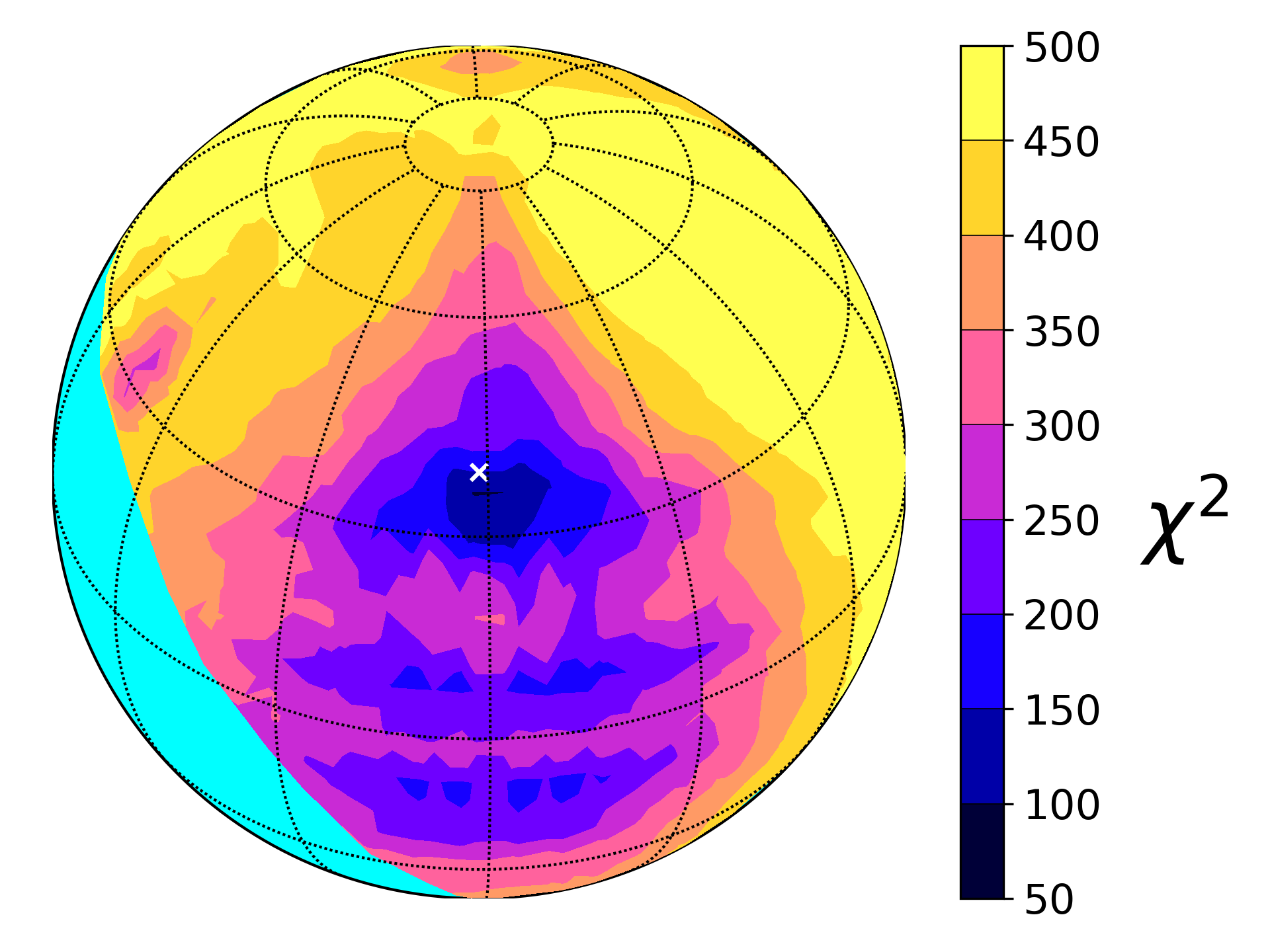}}\hspace{2em}
            %\end{minipage}
            \caption{All-sky $\chi^2$ contour plots for GRB~190117A, as viewed from various directions in different coordinate systems. The top panel shows contour plots in the RA---Dec coordinate system, with the three figures showing views from RA, Dec = (0, $-90$), (0,90), and (113.86, 6.51) from left to right. The three figures in the bottom panel show contours in the satellite $\theta$---$\phi$ coordinates, as viewed from $\theta, \phi$ =(0, 0), (180, 0), and (51.26, 178.51) from left to right. The rightmost figure in each row is a view from the direction of the transient. The darkest region in the contour plot is the most likely location of the GRB from our fits. Where visible, the true location of the GRB is marked with a white cross. }
            \label{fig:all_sky_contour}
\end{figure*}

As discussed in \S\ref{sec:loc_principle}, we calculate the $\chi^2$ for all directions for our GRB sample. We exclude the part of the sky that was occulted by the earth as seen from the satellite at the instant of the burst. Sample localisation contours are shown in Figure~\ref{fig:all_sky_contour}. The left column shows the contours plotted in RA-Dec coordinates, viewed from three different directions in the sky. The right column shows three views of contours plotted in $\theta$--$\phi$ coordinates. Where visible, the location of the GRB is marked with a cross.

Inspecting the contours created for our entire sample, we observe that in most of the cases, the source is well localised with contours showing a single global minimum, while in other cases there are multiple local minima with comparably low $\chi^2$ values. In some of these cases, the actual GRB location is closer to a secondary minimum rather than the global minimum. This suggests that we should select a ``contour filling'' method for describing the GRB location, rather than simply giving an angular offset.

For our 29-GRB sample, we have prior knowledge of the actual source location from other observational instruments. We approximate the GRB position with the HTM grid point closest to it. We denote the $\chi^2$ value at this grid position (cgp) by $\chi^{2}_{\mathrm{cgp}}$. Subsequently, we determine the number of grid points in our analysis that possess $\chi^{2}$ values less than $\chi^{2}_{\mathrm{cgp}}$. The total area with $\chi^2 < \chi^{2}_{\mathrm{cgp}}$ is designated as the ``enclosed area'' $A_{e}$. In Table~\ref{tab:grbinfo}, we provide the obtained values of $A_{e}$ for each GRB in our dataset. Thus, the enclosed area $A_{e}$ is the smallest two-dimensional region, originating from regions with the highest likelihood, that encompasses the true source location.

%=====================================================================================================================================================================

\subsection{Uncertainties in Localisation}
Our localisation method is broadly similar to the source localisation by \fermi-GBM or by the proposed \emph{Daksha} mission. In these cases, it is seen that the uncertainty in the localisation area scales inversely as the detected counts in GRBs detected by GBM~\citep{2020ApJ...893...46V} and simulations for \emph{Daksha} \citep{2022arXiv221112055B}. On the other hand, we can also treat the spacecraft elements as a coded aperture mask being used to obtain source positions. For coded aperture instruments, the localisation position uncertainty scales inversely as SNR \citep{2013ApJS..207...19B}. For bright bursts that we discuss, SNR scales as $\sqrt{\mathrm{counts}}$, and again area uncertainty would scale inversely to the observed counts. Thus, we postulate that the localisation area will scale inversely with the observed counts as,
\begin{equation}
    A_{50}= \frac{K_a}{\rm counts} \label{eq:a50}
\end{equation}
Where $K_a$ is a constant, and $A_{50}$ is defined as the area on the sky that has a 50\% probability of enclosing the true location of the source. %Note that similar inverse relationships could also be defined for any other fractional coverage. 

If we had a large number of GRBs with the same counts then we could have created a distribution of the offsets and calculated $A_{50}$ as the median, thus giving us a value of $K_a$ from Equation~\ref{eq:a50}. However, in practice our GRB sample has different counts for each source, and hence different values for $A_{50}$. To overcome this limitation, we measure the enclosed area $A_e$ for each GRB and normalise it by the expected $A_{50}$ for that source getting a new variable, the ``normalised area'' defined as $a = A_e/ A_{50}$. The distribution of `$a$' gives the probability distribution function of our uncertainties.

\begin{figure}
  \centering
   \includegraphics[width=0.5\textwidth]{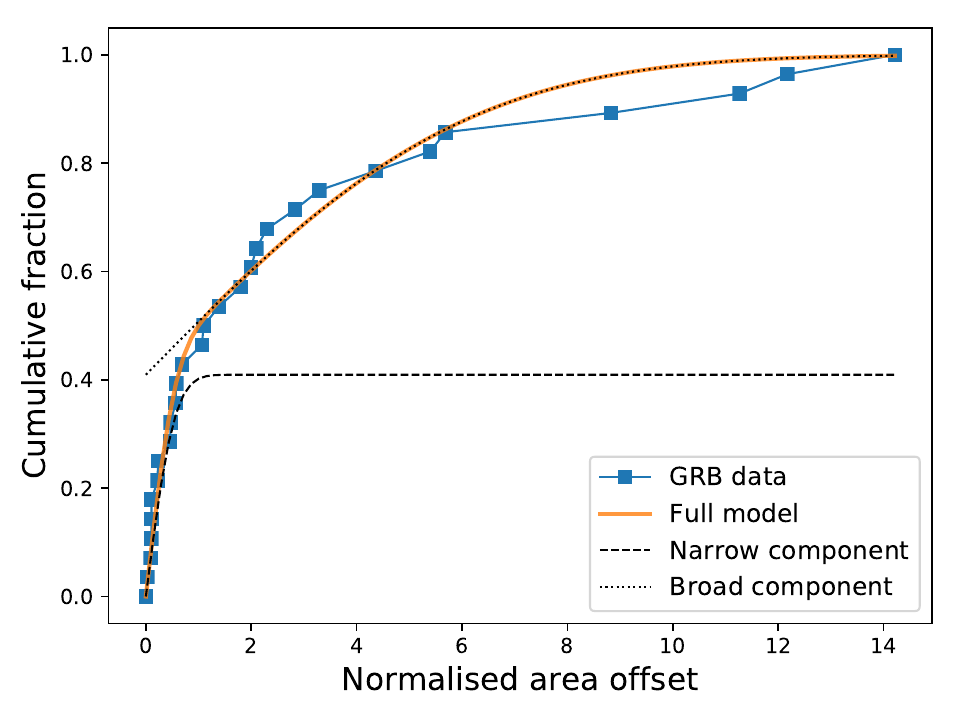}
   \caption{This figure presents the cumulative distribution of normalized area offsets for both GRB data and simulations. The data is modelled using a two-component model comprising a narrow and a broad component, with fitted values of $\sigma_{1} = 0.42$, $\sigma_{2} = 4.76$, and $f = 0.59$, where $f$ represents the fractional contribution of the broad component.}
    \label{fig:cumu}
\end{figure}

To evaluate this underlying probability distribution, we create a cumulative plot of the normalised area (Figure~\ref{fig:cumu}). For normally distributed uncertainties, this should yield an error function (erf). We can see that there is a sharply concentrated core of well-localised GRBs, but we also have a large number of outliers. We find that this is well modelled as a sum of two Gaussians: a narrow component and a broad component with standard deviations $\sigma_1$ and $\sigma_2$ respectively. The corresponding cumulative distribution function is given by Equation~\ref{eq:cdf}, where $f$ is the fractional contribution of the broad component.

\begin{equation}
cdf(a) = \int_{0}^{a} f \frac{1}{\sigma_1\sqrt{2\pi}}~e^{-x^2/2\sigma_{1}^{2}} + (1-f) \frac{1}{\sigma_{2}\sqrt{2\pi}}~e^{-x^2/2\sigma_2^2} dx \label{eq:cdf}
\end{equation}
This distribution gives a good fit to observed data, with $\sigma_{1} =0.42$, $\sigma_{2}=4.76$ and $f=0.59$. We require that the median of this distribution occurs at $a=1$, yielding $K_a = 6.87 \times 10^{6}$. For these distributions, the enclosed probability for various values of $a$ is given in Table~\ref{tab:prob_table}.

We investigated the GRBs to look for any patterns among GRBs that better line up with the narrow versus broad components of the distribution. We find that for GRBs above/below the detector plane ($\theta < 60\degr$ or $\theta> 90\degr$), the $\phi$ values are often correctly measured, and the discrepancy in $\theta$ can be large, even upto 60\degr. Additional minima are usually seen close to the detector plane ($60\degr < \theta < 120\degr$), where the DPHs become relatively featureless. An illustration can be seen in Figure~\ref{fig:all_sky_contour}, where the lower right panel shows two broad local minima above and below the $\theta = 90\degr$ detector plane. However, for this GRB the global minimum happens to be deeper than these spurious local minima, thus giving good localisation. On the other hand, for GRBs close to the detector plane ($60\degr < \theta < 120\degr$), we find that the algorithm correctly places the GRBs within this $\theta$ range, but can have larger errors in both $\theta$ and $\phi$. This effect was also seen in MM21, the relatively featureless nature of the DPH resulted in poorer fits between simulations and observations. The effect can be mitigated to some extent if we know the coarse localisation of the GRBs from other sources --- in which case we can focus only on the relevant local minimum, yielding better localisation. The details of this are not explored further in this work.
\begin{center}
\begin{table}
 \caption{Probability of contained area and corresponding normalised area.\label{tab:prob_table}}
 \centering
    \begin{tabular}{ |c|c| }
    \hline
    Contained probability & Normalised area \\ 
    \hline
    0.10 & 0.12\\
    0.20 & 0.24\\
    0.30 & 0.39\\
    0.40 & 0.59\\
    0.50 &1.00\\
    0.60 & 1.98\\
    0.70 & 3.16\\
    0.80 & 4.56\\
    0.90 & 6.55 \\ 
    \hline
\end{tabular}
\end{table}
\end{center}

\section{Results and Discussions}\label{sec:conclusion}

Under the current counts cutoff criterion of $4000$ counts, we successfully localized $29$ GRBs within the dataset spanning a period of 7 years. The GRBs in our sample have minimum fluence $5.76\times10^{-6}~\mathrm{erg~cm^{-2}~s^{-1}}$and maximum fluence $6.60\times10^{-4}~\mathrm{erg~cm^{-2}~s^{-1}}$. Hence, we can localise GRBs with a minimum fluence of $\sim 6\times10^{-6}~\mathrm{erg~cm^{-2}~s^{-1}}$ to an approximate area $A_{50} \approx$  1700 sq. degree.

\begin{figure}
  \centering
   \includegraphics[width=0.5\textwidth]{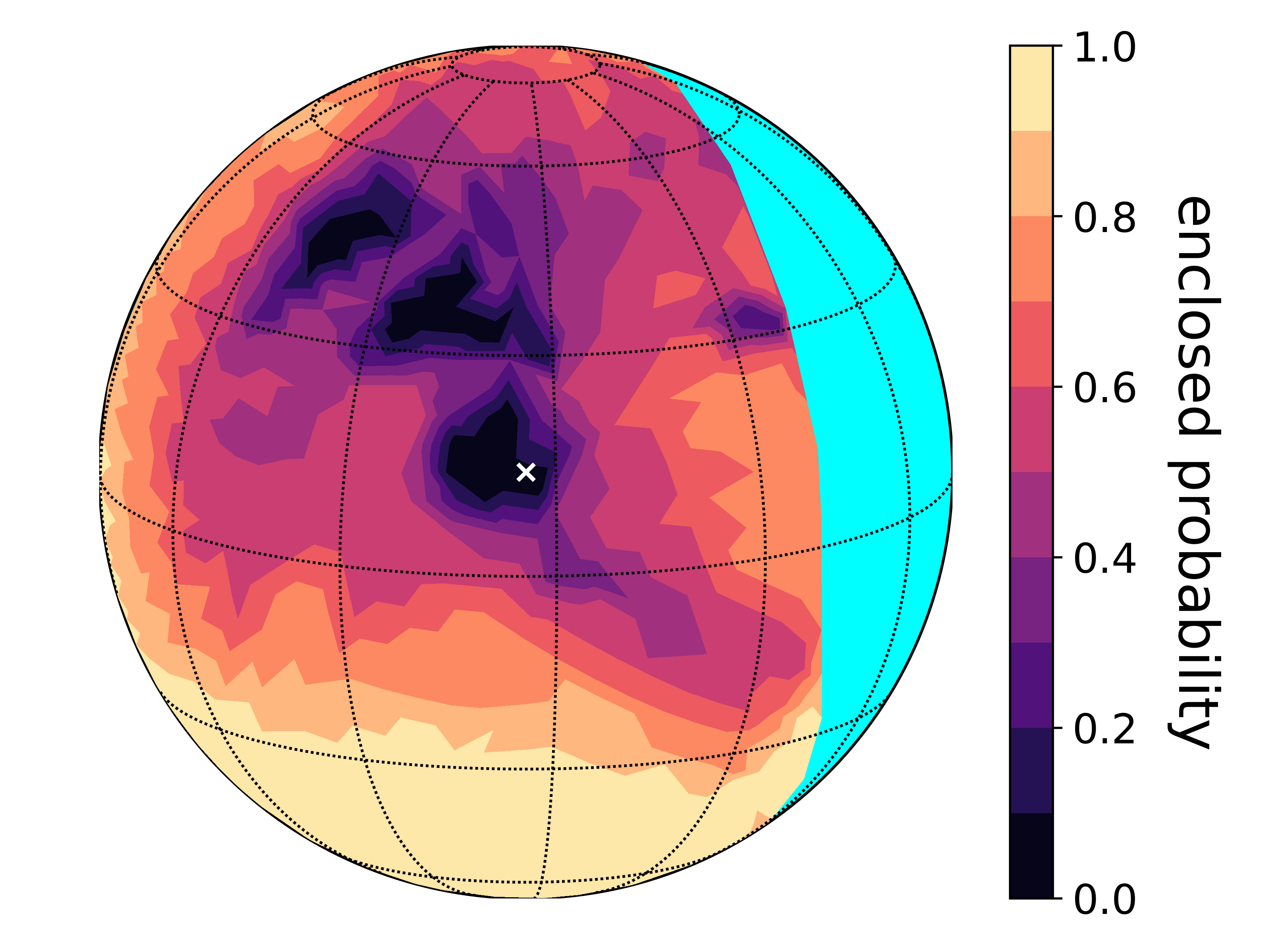}
   \caption{All-sky enclosed probability contour plot for GRB~230817A in RA, Dec; the white cross shows the actual position of the GRB. The cyan-coloured region is the region occulted by the earth.}
    \label{fig:230817a}
\end{figure}

This method is now being used to localize bright GRBs detected by \asat-CZTI. As an illustrative case, consider the recent GRB~230817A detected by CZTI with 3083 counts, which we localized using our current method. The localisation obtained by \fermi for this GRB provided RA = $325.9$, Dec = $14.2$, and an uncertainty radius of $3.3^{\circ}$ \citep{2023GCN.34464....1F}. Our own localisation placed it at RA = $320.35$, Dec = $17.53$, with a corresponding enclosed area $A_{e}=201.43$ square degree. This is well within the expected uncertainty given by Equation~\ref{eq:a50}, ${A_{50}}=2229.7$ square degree. The RA-Dec all-sky probability contour plot for this GRB is depicted in Figure \ref{fig:230817a}.

Our preliminary tests have shown that this method extends well to the localisation of GRBs with fewer counts. This highlights the versatility of our approach, rendering it suitable for localizing GRBs detected by other instruments, particularly in situations where the satellite's structural complexity presents challenges. Currently, this method is being employed to localize two specific GRBs, namely GRB~200503A and GRB~201009A, which will facilitate subsequent polarisation analysis of these events. 

The integration of GRB localisation into the \asat-CZTI GRB pipelines is currently underway, and localisation for bright GRBs will be communicated through future announcements. 

\section*{Acknowledgements}
We thank Prof. A R Rao for useful discussions and guidance. We thank the \asat CZTI team for the help, support and resources.

CZT–Imager is the result of a collaborative effort involving multiple institutes across India. The Tata Institute of Fundamental Research in Mumbai played a central role in spearheading the instrument's design and development. The Vikram Sarabhai Space Centre in Thiruvananthapuram contributed to electronic design, assembly, and testing, while the ISRO Satellite Centre (ISAC) in Bengaluru provided expertise in mechanical design, quality consultation, and project management.
The Inter University Centre for Astronomy and Astrophysics (IUCAA) in Pune was responsible for the Coded Mask design, instrument calibration, and the operation of the Payload Operation Centre. The Space Application Centre (SAC) in Ahmedabad supplied the essential analysis software, and the Physical Research Laboratory (PRL) in Ahmedabad contributed the polarization detection algorithm and conducted ground calibration. Several industries were actively involved in the fabrication process, and the university sector played a crucial role in testing and evaluating the payload.
The Indian Space Research Organisation (ISRO) not only funded the project but also provided essential management and facilitation throughout its development. We also extend our appreciation for the use of the Pegasus High-Performance Computing (HPC) resources at the Inter University Centre for Astronomy and Astrophysics (IUCAA) in Pune. This work utilised various software including Python, AstroPy \citep{astropy}, NumPy \citep{numpy}, Matplotlib \citep{matplotlib}, IDL Astrolib \citep{landsman93}, FTOOLS \citep{blackburn95}, C, and C++. 

%\clearpage
%\appendix
%

%===============================================================================================================================
\begin{landscape}
\begin{table}
%\fontsize{8}{4}
%\footnotesize
\scriptsize
\caption{GRB parameters and computed offsets for the GRBs utilized in Localisation Analysis. Abbreviations: SX - \swift-XRT, SB - \swift-BAT, F - \fermi-GBM, I - IPN, M - Maxi, C - \fermi Catalogue, K - \kw. Spectral parameters are sourced from either \fermi GBM or \kw. \label{tab:grbinfo}}
\renewcommand{\arraystretch}{1.5}
\begin{tabular}{|l|ll|ll|ll|ll|c|c|l|lll|l|}
\hline
%& \multicolumn{15}{|c|}  \\
GRB name & \multicolumn{4}{c|}{Actual} & \multicolumn{4}{c|}{Best fit} & \multicolumn{2}{c|}{Instrument} & \multicolumn{1}{c|}{GRB counts} & \multicolumn{3}{c|}{CZTI localisation}& References\\
\hline
& RA& Dec& $\theta$& $\phi$& RA& Dec& $\theta$& $\phi$& localisation& spectra &  & ${A_{e}}$& ${A_{50}}$& a & (localisation,spectrum)\\
\hline
GRB~180809B &299.69&-15.29&26.45&198.88&3.92&36.66&106.08&202.99& SX & K&19662&201.43&349.61&0.57& \cite{liu2018grb},\cite{2018GCN.23128....1S} \\
GRB~170527A &195.19&0.94&26.53&101.54&204.91&-4.94&27.8&126.43& F & K  &13290&1188.43&517.22&2.29& \cite{stanbro2017grb}, \cite{2017GCN.21166....1F}\\
GRB~200311A &204&-49.7&30.91&157.64&204.87&-45.06&35.34&154.89& F & K  &6102&120.85&1126.40&0.10& \cite{2020GCN.27363....1F}, \cite{2020GCN.27398....1R}\\
GRB~200412A &140.01&-41.67&36.15&270.08&107.91&20.68&105.14&266.03& F & F &6176&100.71&1113.05&0.09& \cite{kunzweiler2020grb}, \cite{2020GCN.27550....1M}\\
GRB~180914A &53.08&-5.62&40.67&216.51&353.55&-49.83&106.66&225& SX & K &13567&705&506.67&1.39& \cite{d2018grb}, \cite{2018GCN.23254....1T}\\
GRB~180427A &283.33&70.3&40.81&257.79&294.05&5.93&105.14&266.03& F & F &8366&564&303.12&1.86& \cite{bissaldi2018grb},\cite{2018GCN.22678....1B}\\
GRB~190117A &113.86&6.51&51.26&178.51&115.73&8.69&54&177.48& F & K&6038&0&1138.44&0& \cite{hurley2019ipn17}, \cite{2019GCN.23782....1F}\\
GRB~160802A &35.29&72.69&52.95&273.11&20.67&71.47&54&267.48& F & F &12255&60.42&560.89&0.10& \cite{bissaldi2016grb}, \cite{2016GCN.19754....1B}\\
GRB~180806A &11.55&24.33&54.69&264.49&12.39&26.73&54&267.48& SX & F &4191&100.71&1639.94&0.06& \cite{gibson2018grb}, \cite{2018GCN.23095....1M}\\
GRB~190731A &339.72&-76.57&57.44&126.29&113.1&-53.72&55.62&69.97& SX & F &9187&3263.17&748.21&4.36& \cite{2019GCN.25240....1F}, \cite{2019GCN.25248....1R}\\
GRB~190928A &36.57&29.46&57.68&231.25&30.79&26.37&54.74&225& I & K &24175&302.14&284.34&1.06& \cite{hurley2019ipn28}, \cite{2019GCN.25868....1F}\\
GRB~221209A &241.7&48.1&58.4&310.81&216.85&7.15&104.46&310.55& F & F &4598&705&1494.97&0.47& \cite{2022GCN.33031....1F}, \cite{2022GCN.33034....1B}\\
\hline
GRB~160910A &221.44&39.06&65.53&333.46&105.72&63.45&105.42&280.06& SX & F &5315&7352.21&1293.16&5.68& \cite{veres2016grb}, \cite{2016GCN.19901....1V}\\
GRB~220408B &94.8&-50.9&73.03&275.64&88.26&-49.23&74.58&280.06& F & F &4838&40.28&1420.83&0.03& \cite{2022GCN.31854....1F}, \cite{2022GCN.31906....1B}\\
GRB~210204A &117.08&11.4&75.28&166.59&145.81&3.88&64.12&195.73& SX & F&7859&201.43&874.69&0.23& \cite{2021GCN.29232....1F}, \cite{2021GCN.29393....1B}\\
GRB~230204B &197.64&-21.71&76.01&253.9&245.03&-43.73&99.98&292& SX & F &15725&2356.73&437.15&5.39& \cite{d2023grb}, \cite{2023GCN.33288....1P}\\
GRB~221022B &167.4&15.4&77.19&312.96&104.01&41.8&93.83&254.83& F & F&6755&12387.97&1017.54&12.17& \cite{2022GCN.32820....1F}, \cite{2022GCN.32830....1P} \\
GRB~181222A &270.73&-38.84&79.78&346.37&274.59&-31.9&86.2&350.53& M & F &6205&3645.89&1107.84&3.29& \cite{sakamaki2018grb}, \cite{2018GCN.23548....1V}\\
GRB~210822A &304.43&5.27&84.75&168.11&206.6&-1.9&56.73&65.01& SX & K &5504&1369.72&1248.89&1.09& \cite{page2021grb}, \cite{2021GCN.30694....1F}\\
GRB~201216C &16.37&16.51&88.6&116.81&19.83&17.87&87.98&120.34& SX & F&7212&100.71&953.16&0.10& \cite{beardmore2020grb}, \cite{2020GCN.29073....1M}\\
GRB~170607B &257.1&-35.7&96.33&178.63&257.48&-8.27&70.63&188.36& F & F&4703&3061.74&1461.37&2.09& \cite{hamburg2017grb}, \cite{2017GCN.21238....1H}\\
GRB~160106A &181.6&17.5&106.09&255.65&165.47&-0.19&129.84&256.61& C & F &4229&2920.74&1625.21&1.79& \cite{vonKienlin2020}, \cite{vonKienlin2020}\\
\hline
GRB~160607A &13.66&-4.94&138.84&315.76&343.28&42.15&91.91&283.2& SX & F&4867&2820.02&1412.24&1.99& \cite{ukwatta2016grb}, \cite{2016GCN.19511....1T}\\
GRB~160623A &315.29&42.22&140.46&118.06&20.18&-7.38&74.13&73.49& SX & K &11414&5317.76&602.23&8.83& \cite{mingo2016grb}, \cite{2016GCN.19554....1F}\\
GRB~171010A &66.58&-10.46&142.51&242.01&71.72&-6.31&137.37&248.36& SX & F &39144&80.57&175.61&0.45& \cite{d2017grb}, \cite{2017GCN.21992....1P}\\
GRB~170906A &203.99&-47.12&146.24&263.16&219.49&-61.73&132.64&247.06& SX &F&5812&664.72&1182.71&0.56& \cite{siegel2017grb}, \cite{2017GCN.21839....1H}\\
GRB~171227A &280.7&-35&146.48&353.57&228.7&37.56&59.68&2.34& F & F &7367&10514.67&933.03&11.26& \cite{hui2017grb}, \cite{2017GCN.22289....1H}\\
GRB~190530A &120.53&35.47&154.5&79.86&259.8&79.62&91.91&76.8& SX & F &11630&8399.65&591.053&14.21& \cite{biltzinger2019fermi}, \cite{2019GCN.24692....1B}\\
GRB~160821A &171.24&42.34&156.17&59.26&23.07&61.15&88.11&97.52& SB & F &21459&906.43&320.34&2.82& \cite{siegel2016grb}, \cite{2016GCN.19835....1S}\\
\hline
\end{tabular}
\end{table}
\end{landscape}

%================================================================================================================================================
\section*{Data Availability}
The data underlying this article will be shared on reasonable request to the corresponding author.

\vspace{-1em}
%\bibliographystyle{apj}
\bibliography{main_bibfile}
%======================================================================
\begin{comment}
  \appendix

\section{Localisation uncertainty}\label{sec:app}
Consider a simplified case of a detector in the XY plane, with a shadow-casting element located at a height $h$ above it.  
\end{comment}

%=====================================================================================================================================
%\onecolumn
%\section{Impact of scaling correction}
%%%%%%%%%%%%%%%%%%%%%%%%%%%%%%%%%%%%%%%%%%%%%%%%%%

% Don't change these lines
\bsp	% typesetting comment
\label{lastpage}
\end{document}